\documentclass[conference,twocolunm,10pt]{IEEEtran}
\usepackage{epsfig,amssymb,amsmath}
\usepackage{colortbl}     
\begin{document}

\title{\huge{Interference Alignment Through User Cooperation\\
for Two-cell MIMO Interfering Broadcast Channels}}
\author{\authorblockN{
Wonjae Shin, Namyoon Lee, Jong-Bu Lim, Changyong Shin, and Kyunghun Jang}\\
\authorblockA{
Communication Laboratory,
 Samsung Advanced Institute of Technology\\ Samsung Electronics Co.,
 Ltd.,
Yongin-Si, Gyeonggi-Do, Korea 446-712
\\
    Email: \{wonjae.shin, namyoon.lee, jongbu.lim, c.y.shin, khjang\}@samsung.com\\
}} \maketitle \thispagestyle{empty}
\begin{abstract}



This paper focuses on two-cell multiple-input multiple-output (MIMO) Gaussian interfering broadcast channels (MIMO-IFBC) with $K$ cooperating users on the cell-boundary of each BS. It corresponds to a downlink scenario for cellular networks with two base stations (BSs), and $K$ users equipped with Wi-Fi interfaces enabling to cooperate among users on a peer-to-peer basis. In this scenario, we propose a novel interference alignment (IA) technique exploiting user cooperation. Our proposed algorithm obtains the achievable degrees of freedom (DoF) of $2K$ when each BS and user have $M=K+1$ transmit antennas and $N=K$ receive antennas, respectively. Furthermore, the algorithm requires only a small amount of channel feedback information with the aid of the user cooperation channels. The simulations demonstrate that not only are the analytical results valid, but the achievable DoF of our proposed algorithm also outperforms those of conventional techniques.
\end{abstract}
\begin{keywords}
Interference Alignment, User Cooperation, MIMO Interfering Broadcast Channels $\vspace{1mm}$
\end{keywords}


\section{Introduction}

To increase system capacity for next generation cellular systems (e.g., \cite{3GPP}, \cite{16x}), interference mitigation methods such as network multiple-input multiple-output (MIMO) and coordinated multipoint transmission and reception (CoMP) have been actively discussed for the multi-cell and multi-user downlink transmission. Specifically, considering the case that each base station (BS) sends two independent messages to two different users in the cell it covers, respectively, we note that each user suffers from the inter-user interference (IUI) and the inter-cell interference (ICI).  To mitigate those interferences effectively in the case, a simple coordinated zero-forcing (ZF) scheme was proposed in \cite{Park}. This scheme reduces both ICI and IUI simultaneously in the multiple-input single-output interfering broadcast channels (MISO-IFBC). Since this scheme considered a user with a single antenna, it only designed the transmit beamforming vectors to remove both IUI and ICI. In addition, it was shown that the coordinated ZF scheme achieves the optimal degrees of freedom (DoF) on that channel through the converse argument. In \cite{Kim}, the coordinated ZF scheme for the MIMO-IFBC was extended by taking multiple receive antennas into account. Furthermore, it was shown in \cite{Kim} that when there are a BS with $M$ transmit antennas and $K$ users with $N$ receive antennas in each cell, the coordinated ZF scheme achieves the DoF of $\mathrm{min}\{2M,2KN,\mathrm{max}(M,N)\}$ in the two-cell MIMO-IFBC.
In \cite{Tse} and \cite{Suh}, a precoding scheme called the subspace interference alignment (IA) for the MIMO-IFBC was introduced by appropriately exploiting the concept of the original IA in \cite{Cadambe}-\cite{MMK}. The subspace IA is composed of two cascaded precoders. The first precoder is to put ICI vectors on a multi-dimensional subspace. The second procoder is constructed so that the IUI vectors lie on the subspace spanned by the ICI vectors. Consequently, the signal space occupied by the IUI vectors is aligned with the signal space spanned by the ICI vectors, which increases the subspace dimension for desired signal vectors especially in the symmetric antenna case. In addition, it is noted that this technique requires the same amount of feedback as a conventional single cell multi-user MIMO system.

\begin{figure}
\centering
\includegraphics[width=3in]{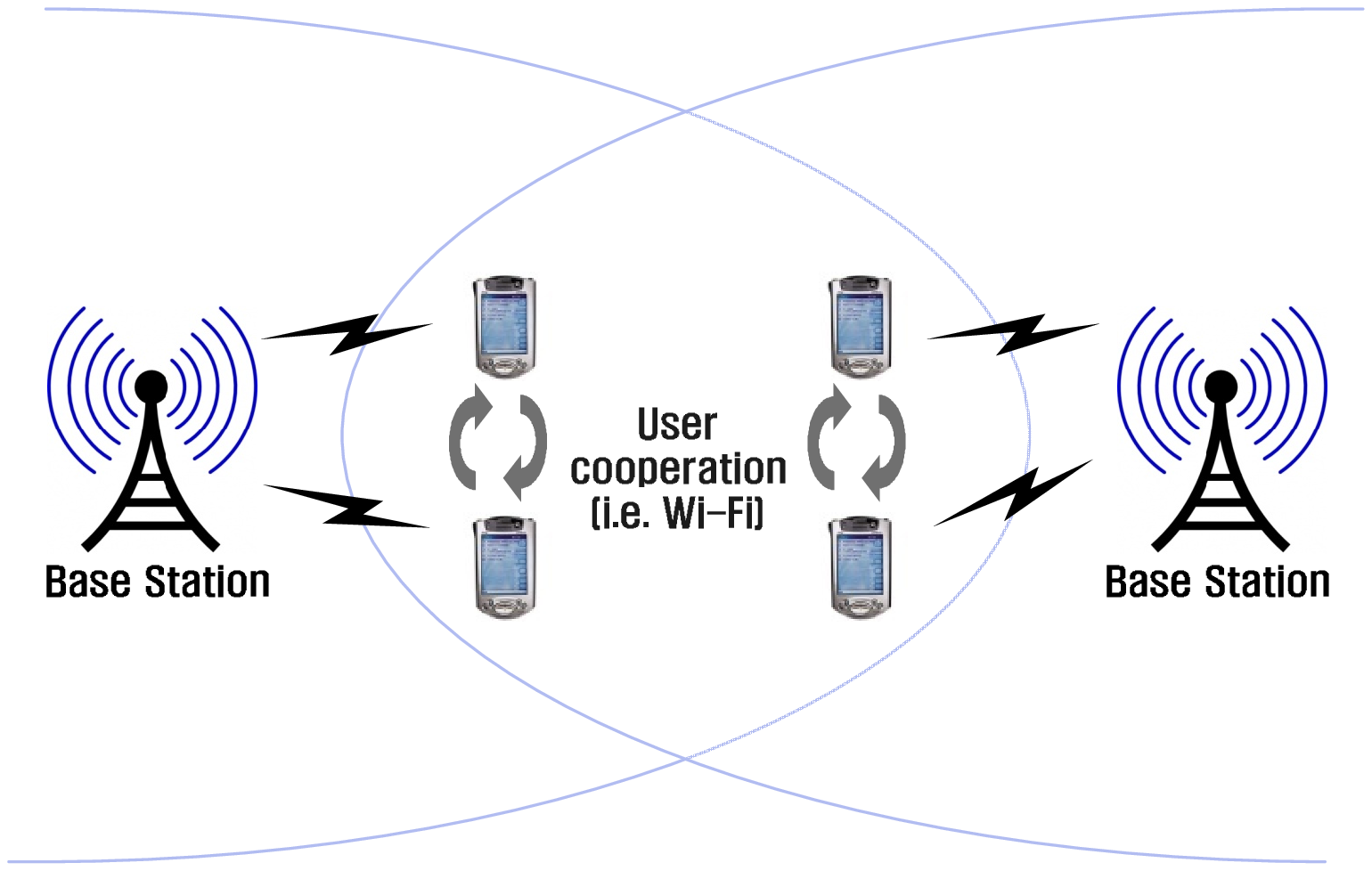}
\centering \caption{Two-cell MIMO Interfering Broadcast Channels
(MIMO-IFBC) with cooperating users.}
\end{figure}

\begin{figure*}[pt!]
\begin{eqnarray}
&&R^{[k,i]}({\bf v}^{[1,1]}, {\bf v}^{[2,1]}, \ldots, {\bf v}^{[K,1]}, {\bf v}^{[1,2]},
{\bf v}^{[2,2]}, \ldots, {\bf v}^{[K,2]}, {\bf w}^{[k,i]})\nonumber\\
&&=\mathrm{log}_2\left(1+\frac{\left|{\bf
w}^{[k,i]\dag}{\bf H}^{[k,i]}_{i}{\bf
v}^{[k,i]}\right|^2}{\sigma^2+\sum_{k'=1,k'\neq
k}^{K}\left|{\bf w}^{[k,i]\dag}{\bf H}^{[k,i]}_{i}{\bf
v}^{[\bar{k^{\prime}},i]}\right|^2
+\sum_{k^{\prime}=1}^{K}\left|{\bf
w}^{[k,i]\dag}{\bf H}^{[k,i]}_{\bar{i}}{\bf
v}^{[k^{\prime},\bar{i}]}\right|^2}\right).
\end{eqnarray}
\hrulefill
\end{figure*}

Furthermore, since the Wi-Fi and the cellular network are rapidly converging for future wireless communication systems \cite{16x}, it will be possible to cooperate among users in cellular systems by exploiting Wi-Fi links. For instance, an upcoming specification from the Wi-Fi Alliance, the Wi-Fi peer-to-peer \cite{WiFi}, will support a new way for user terminals equipped with the Wi-Fi to enable to make direct connections to one another quickly and conveniently.
In general, collaboration levels of user cooperative links could be classified into two categories: received-signal forwarding and channel state information (CSI) sharing scenarios. Most prior works on user cooperation have considered the received-signal forwarding cases, which act as a cooperative relay in different channel models such as broadcast channel \cite{Dabora}-\cite{Kwon}, two-user interference channel \cite{Host-Madsen}-\cite{Wang}, and one-sided interference channel \cite{Yu} etc.


In this paper, we propose a novel cross-network interference alignment algorithm exploiting another channel resource available for user cooperation, i.e., a cooperation channel among cell-edge users for the two-cell MIMO-IFBC. Our algorithm jointly designs transmit and receive beamforming vectors for the two-cell MIMO-IFBC using CSI obtained by user cooperation via the Wi-Fi networks. In terms of the achievable DoF and the amount of channel feedback, we also investigate the benefits of our algorithm offered by the utilization of the Wi-Fi links to share CSI among users.

The remainder of this paper is organized as follows. Section II describes the system model considered in this paper. In Section III, we present our proposed algorithm by a simple example and derive the main result of this paper. In Section IV, we demonstrate that our proposed algorithm outperforms than the conventional techniques via simulation results. Finally, the paper is concluded in Section VII.\\

\section{System model} \label{sec:System Model}
In this section, we describe a system model for the two-cell
MIMO-IFBC as shown in Fig. 1. The system of the two-cell MIMO-IFBC consists of two BSs with $M$ antennas per BS and $K$ users with $N$ receive antennas per user in each cell. It is assumed that $K$ users in each cell can cooperate to share their CSI using the Wi-Fi links which offers wireless connectivity on a peer-to-peer basis without the assistance of wired network. This assumption is valid in the case where cooperating users are located in a local area, such as in a campus or a library.
 For notation convenience, we refer to the $k$-th user in the $i$-th cell as user $[k,i]$. We assume each BS tries to convey one data stream per user to its corresponding users.

\subsection{BS-User Links for Cellular Networks}
The received signal at the user $[k,i]$ is represented as
\begin{eqnarray}
&&\hspace{-8mm}{\bf y}^{[k,i]}=\sum_{i^{\prime}=1}^{2} {\bf H}^{[k,i]}_{i^{\prime}} \sum_{{k}^{\prime}=1}^{K} {\bf v}^{[k^{\prime},i]}{s}^{[k^{\prime},i]}+{\bf n}^{[k,i]}\nonumber\\
&&=\underbrace{{\bf H}^{[k,i]}_{i}{\bf v}^{[k,i]}{s}^{[k,i]}}_{\textrm{desired signal}}
               +\sum_{{k}^{\prime}=1, {k}^{\prime}\neq k}^{K}\underbrace{{\bf H}^{[k,i]}_{i}{\bf v}^{[k^{\prime},i]}{s}^{[k^{\prime},i]}}_{\textrm{\textrm{inter-user interference}}}\nonumber\\
               &&\hspace{5mm}+\sum_{k^{\prime}=1}^{K}\underbrace{{\bf H}^{[k,i]}_{\bar{i}}{\bf v}^{[k^{\prime},\bar{i}]}{s}^{[k^{\prime},\bar{i}]}}_{\textrm{inter-cell interference}}+{\bf{n}}^{[k,i]},
\end{eqnarray}
where $s^{[k,i]}$ denotes the transmitted symbol for the $k$-th user in the $i$-th cell, satisfying an average power constraint, $\mathbb{E}\left[ \sum_{k=1}^{K}\left\|{\bf v}^{[k,i]}{s}^{[k,i]}\right\|^2 \right]\leq P$, and ${\bf v}^{[k,i]}$ is the linear beamforming vector for carrying the symbol, $s^{[k,i]}$, with unit norm constraint, i.e., $\left\|{\bf v}^{[k,i]}\right\|=1$. In (2), ${\bf n}^{[k,i]}$ is the $N \times 1$ additive white
Gaussian noise (AWGN) vector with variance $\sigma^2$ per entry at the $k$-th user in the $i$-th cell, and ${\bf H}_{j}^{[k,i]}$ is the $N \times M$ channel matrix from the BS $j$ to the user
$[k,i]$. We define $\bar{i}$ as the cell index which is not equal to $i$, and
$\forall i,\bar{i}\in{\{1,2\}}$. The channel matrices
${\bf H}_{j}^{[k,i]}$ for $\forall i,j\in{\{1,2\}}$ and $\forall k
\in \{1,2,\ldots,K\}$ are generated so that each entry of the matrix is independent and identically distributed (i.i.d.) according to $\mathcal{CN}(0,1)$,
 and it is assumed that each channel obeys a frequency flat block-fading channel model.
 We also suppose that perfect channel state information (CSI) is available at all users.
 Each user decodes the desired signals coming from its corresponding BS by multiplying the receive beamforming vector,
  and the signal at the user $[k,i]$ after receiver combining is given by
\begin{eqnarray}
{\bf \tilde{y}}^{[k,i]}&=&{\bf w}^{[k,i]\dag}{{\bf H}^{[k,i]}_{i}{\bf v}^{[k,i]}{s}^{[k,i]}}\nonumber\\&+&{\bf w}^{[k,i]\dag}{\sum_{{k}^{\prime}=1, {k}^{\prime}\neq k}^{K}{\bf H}^{[k,i]}_{i}{\bf v}^{[k^{\prime},i]}{s}^{[k^{\prime},i]}}\nonumber\\
               &+&{\bf w}^{[k,i]\dag}\sum_{k^{\prime}=1}^{K}{\bf H}^{[k,i]}_{\bar{i}}{\bf v}^{[k^{\prime},\bar{i}]}{s}^{[k^{\prime},\bar{i}]}
               +{\bf \tilde{n}}^{[k,i]},
\end{eqnarray}
where ${\bf w}^{[k,i]}$ denotes the
receive beamforming vector for the user $[k,i]$, whose size is $N\times 1$, and $ {\bf \tilde{n}}^{[k,i]}={\bf
w}^{[k,i]\dag}{\bf n}^{[k,i]}$ is the effective noise vector,
each entry of which is distributed as $\mathcal{CN}(0,1)$. The notation $\left(\cdot\right)^\dag$ means the conjugate transpose operator.

\subsection{User-Cooperative Links for Wi-Fi Networks}
The user-cooperative links are connected by the Wi-Fi networks in an ad hoc mode \cite{WiFi} without any adverse impacts or interference caused by the cellular network. In this paper, these links are assumed to be noiseless with achievable rates which are enough to share CSI of interfering channels among cooperating users in each one cell.

\subsection{Degrees of Freedom (DoF)}
For the given set of beamforming vectors, ${\bf v}^{[k,i]}$ and ${\bf w}^{[k,i]}$, where $\forall k
\in \{1,2,\ldots,K\}$ and $\forall i
\in \{1,2\}$, the achievable rate at the user $[k,i]$ supported by the BS $i$ is written as given in equation (1).

We define the rate region $\mathfrak{R}$ to be the set of all
rates that can be achieved by using the transmit and receive
beamforming vectors under individual power constraints, which
is given by
\begin{eqnarray}
\mathfrak{R}\triangleq \{ \left(R^{[1,1]}, R^{[2,1]}, \ldots, R^{[K,1]}, R^{[1,2]},
R^{[2,2]}, \ldots, R^{[K,2]}\right)\nonumber \\
\in {\mathbb{R}}^{2K}_{+} |R^{[k,i]}
\mathrm{\,\,as \,\,in\,\, }(1), \forall i,k \}.
\end{eqnarray}
Further, we define the degrees of freedom which is the
pre-log factor of the sum rate. This is one of the key metrics for assessing the performance of the system in the
multiple antenna systems at the high SNR regime, which is defined as
\begin{eqnarray}
d\triangleq
\lim_{\rho\rightarrow\infty}\frac{R(\rho)}{\mathrm{log}(\rho)}=\sum_{k\in\{1,\ldots,K\}}\sum_{i\in\{1,2\}}d^{[k,i]},
%
\end{eqnarray}
where
$R(\rho)=\sum_{k\in\{1,\ldots,K\}}\sum_{i\in\{1,2\}}R^{[k,i]}(\rho)$
denotes the sum rate at signal-to-noise ratio
$(\rho=\frac{P}{\sigma^2})$.\\

\section{New IA scheme through user cooperation}\label{sec:Inner Bound}

In this section, we introduce a new IA scheme which mitigates both ICI and IUI simultaneously in the two-cell MIMO-IFBC with help of user cooperation, and investigate the benefits of user cooperation compared with existing schemes in terms of DoF and the amount of channel feedback.

\subsection{Motivating example for $(M,N,K)=(3,2,2)$}
To explain our scheme and its benefits clearly, we start with a simple case of $(M,N,K)=(3,2,2)$ as shown in Fig. 2.
The BS $1$ wants to deliver two symbols, $s^{[1,1]}$ and $s^{[2,1]}$, to the user $[1,1]$ and user $[2,1]$ using the transmit beamforming vectors ${\bf v}^{[1,1]}$ and ${\bf v}^{[2,1]}$, respectively. In general, for given receive beamforming vectors, the minimum number of transmit antennas is 4 so that the transmit beamforming vectors cancel out all ICI and IUI. For example, in order to transmit the symbol $s^{[1,1]}$ without causing any interference to the other users, the beamforming vector ${\bf v}^{[1,1]}$ should satisfy the following condition,
\begin{eqnarray}
&&\hspace{-11mm}{\bf v}^{[1,1]}\subset\mathrm{null}\Bigg(\bigg[\underbrace{{(\bf w}^{[2,1]\dag}{\bf H}_{1}^{[2,1]})^{\dag}}_{\mathrm{effective\,\,IUI\,\,channel}},\nonumber\\
&&\hspace{12mm}\underbrace{({\bf w}^{[1,2]\dag}{\bf H}_{1}^{[1,2]})^{\dag},
\,\,\,\,\,\,  ({\bf w}^{[2,2]\dag}{\bf
H}_{1}^{[2,2]})^{\dag}}_{\mathrm{effective\,\,ICI\,\,channels}}
\bigg]^{\dag}\Bigg),
\end{eqnarray}
where $\mathrm{null}(\cdot)$ denotes an orthonormal basis for the null space of a matrix. However, our proposed scheme can remove both ICI and IUI with 3 transmit antennas by performing ICI channel alignment. In the following four steps, we present our transmit and receive beamfoming design method enabling the ICI channel alignment through user cooperation without the need of global CSI.\\

\textit{\textbf{Step 1:}} \textbf{Sharing interfering channels among users}

In order to share CSI between co-located cell-edge users, user terminals perform user cooperation using the Wi-Fi network in ad-hoc mode. To be specific, the user $[1,2]$ and the user $[2,2]$ exchange CSI of interfering channels to cooperatively design receive beamforming vectors.\\


\textit{\textbf{Step 2:}} \textbf{Designing the receive
beamforming vectors}

By using CSI of interfering channels acquired in the \textit{ step 1}, the user $[1,2]$ and user $[2,2]$ design the receive beamforming vectors ${\bf w}^{[1,2]}$ and ${\bf w}^{[2,2]}$, so that the effective ICI channels from the BS 1 are aligned with each other, which is
\begin{eqnarray}
\mathrm{span} \left( {\bf H}_{1}^{[1,2]\dag}{\bf w}^{[1,2]}
\right) =\mathrm{span} \left( {\bf H}_{1}^{[2,2]\dag}{\bf
w}^{[2,2]} \right),
\end{eqnarray}
where $\mathrm{span}(\cdot)$ denotes the space spanned by the column vectors of a matrix.\\
We can find out the intersection subspace satisfying the condition (7) above by solving the following matrix equation,
\begin{eqnarray}
\underbrace{\left[%
\begin{array}{cccc}
  {\bf I}_{M} & -{\bf H}_{1}^{[1,2]\dag} & {\bf 0} \\
  {\bf I}_{M} & {\bf 0} & -{\bf H}_{1}^{[2,2]\dag} \\
\end{array}%
\right]}_{6\times 7}
\left[%
\begin{array}{cccc}
  {\bf h}_{1}^{\mathcal{ICI}} \\
  {\bf w}^{[1,2]} \\
  {\bf w}^{[2,2]} \\
\end{array}%
\right]={\bf M}_1{\bf x}_1={\bf 0},
\end{eqnarray}
where ${\bf h}^{\mathcal{ICI}}_{1}$ implies the direction of aligned effective interference channels from the BS $1$ to the user $[1,2]$ and user $[2,2]$ after applying the receiver beamforming vectors. Since the size of the matrix ${\bf M}_1$ is $6 \times 7$, it has one dimensional null space. Therefore, the receive beamforming vectors for ICI channel alignment can be obtained explicitly with probability one.\\

\textit{\textbf{Step 3:}} \textbf{Feedback the effective channels to the BS}

Each user feeds back equivalent channels after applying the receive beamforming vectors determined in the \textit{step 2} instead of channels itself through uplink feedback channels for the cellular networks to its corresponding BS. To be specific, the user $[1,2]$ is required to feed back both the effective serving and interfering channel vectors after applying the receive beamforming vectors, i.e., ${\bf w}^{[1,2]\dag}{\bf H}^{[1,2]}_2\in\mathbb{C}^{1\times 3}$ and ${\bf w}^{[1,2]\dag}{\bf H}^{[1,2]}_1\in\mathbb{C}^{1\times 3}$.\\

\begin{figure}
\centering
\includegraphics[width=3in]{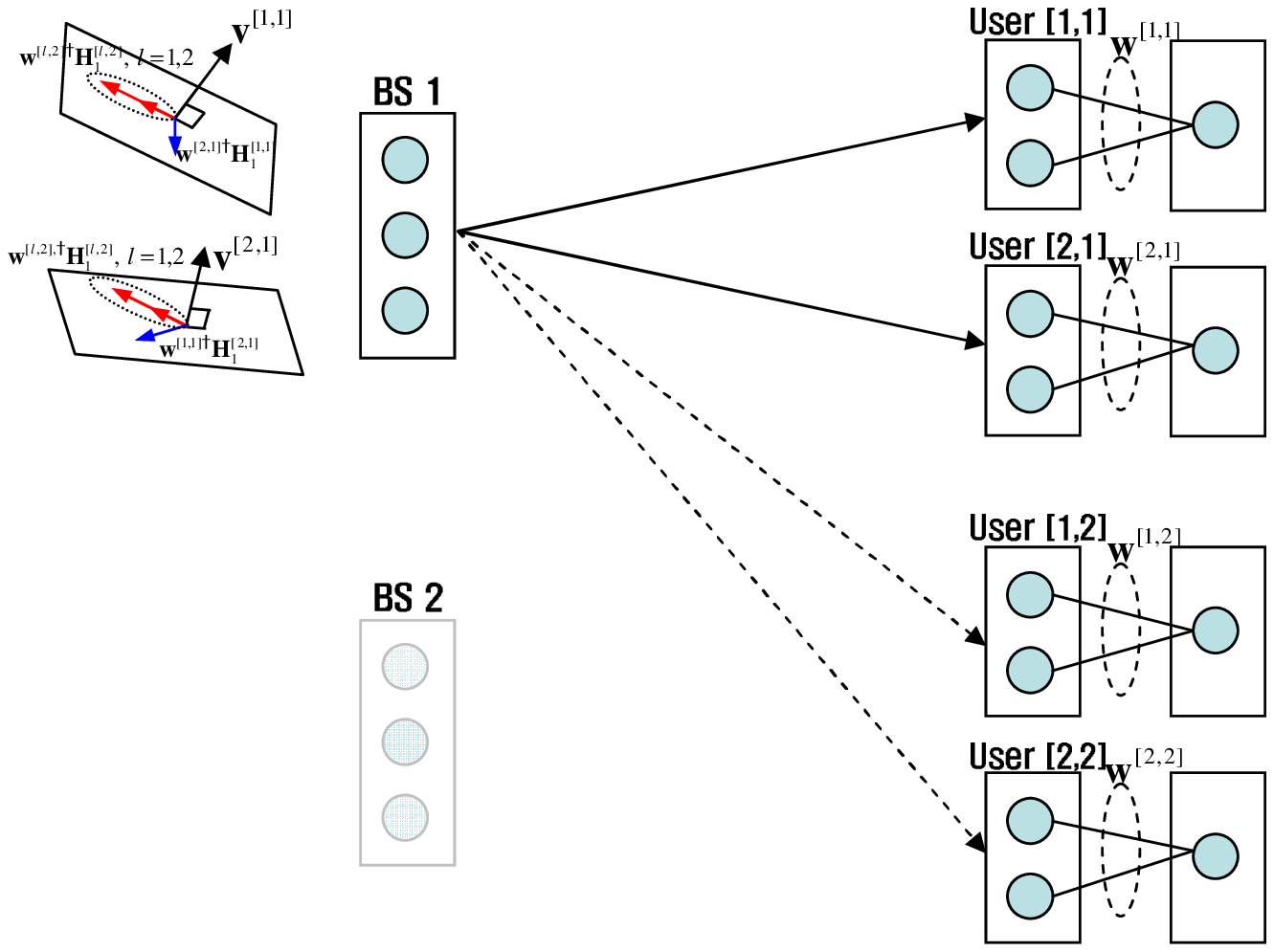}
\centering \caption{The concept of the proposed IA scheme for the (3,2,2) MIMO-IFBC.}
\end{figure}
\textit{\textbf{Step 4:}} \textbf{Choosing the transmit
beamforming vectors}

Since the effective ICI channels are aligned with each other,
the BS 1 can consider two different ICI channel vectors as a one ICI channel vector which spans one dimensional subspace as shown in Fig. 2. Therefore, if the beamforming vectors ${\bf v}^{[1,1]}$ and ${\bf v}^{[2,1]}$ are designed with the effective channel information in the\textit{ step 3} as
\begin{eqnarray}
{\bf v}^{[1,1]} \subset \mathrm{null}\left(\left[ ({\bf
w}^{[2,1]\dag}{\bf H}_{1}^{[2,1]})^{\dag},  \,\,\, {\bf
h}^{\mathcal{ICI}}_{2} \right]^{\dag}\right),\nonumber \\  {\bf
v}^{[2,1]}\subset \mathrm{null}\left(\left[ ({\bf
w}^{[1,1]\dag}{\bf H}_{1}^{[1,1]})^{\dag},  \,\,\, {\bf
h}^\mathcal{{ICI}}_{2} \right]^{\dag}\right),
\end{eqnarray}
the BS 1 can send the symbols $s^{[1,1]}$ and $s^{[2,1]}$ to the user $[1,1]$ and user $[2,1]$ without any interferences, respectively.





\subsection{General case for multiple user $K>2$}
We now generalize the proposed IA scheme explained in previous section for multiple cell-edge users in each cell, $K>2$. The following theorem is the main result of this paper.




$\mathrm{\textit{\textbf{Theorem 1}}}$: For the two-cell MIMO-IFBC where
each BS supports $K$ users simultaneously, we
can achieve $2K$ degrees of freedom if the BS and user have $M=K+1$ and $N=K$ antennas, respectively.

$\mathrm{\textit{Proof )}} \,\,\,$ The achievability
proof is provided by using an extension of the interference
alignment scheme presented in previous subsection. 
To prove Theorem 1, we show that $d^{[k,i]}=1, \forall k\in\{1,2,...K\}, \forall i\in\{1,2\}$, is achieved when the numbers of transmit antennas $M$ and receive antennas $N$ are equal to $K+1$ and $K$, respectively.
To be able to decode a single stream at each user, the interferences coming from $\bar{i}$-th cell and $i$-th cell should be aligned within no more than $N-1$ dimensional in order to obtain 1 interference free dimension from a $N$-dimensional received signal vector. Without loss of generality, the transmit and receive beamforming vectors to mitigate both ICI and IUI simultaneously in the two-cell MIMO-IFBC can be designed as in equation (10) on the top of the next page.
\begin{figure*}[t]
\begin{eqnarray}
&&\hspace{-10mm}\mathrm{span}({\bf v}^{[k,i]}) \subset \mathrm{null}\bigg(\bigg[
\underbrace{({\bf
w}^{[1,\bar{i}]\dag}{\bf H}^{[1,\vec{i}]}_{i})^{\dag},\,\cdots, ({\bf
w}^{[K,\bar{i}]\dag}{\bf
H}^{[K,\vec{i}]}_{i})^{\dag}}_{\mathrm{effective \,\, ICI\,\, channels}}\nonumber\\
&&\hspace{20mm}
\underbrace{({\bf w}^{[1,i]\dag}{\bf
H}^{[1,i]}_i)^{\dag},\cdots,({\bf w}^{[k-1,i]\dag}{\bf
H}^{[k-1,i]}_i)^{\dag},({\bf w}^{[k+1,i]\dag}{\bf
H}^{[k+1,i]}_i)^{\dag},\cdots,({\bf w}^{[K,i]\dag}{\bf
H}^{[K,i]}_i)^{\dag}}_{\mathrm{effective
\,\,IUI\,\,channels}}
 \bigg]^{\dag}\bigg).
\end{eqnarray}
\hrulefill
\end{figure*}

\begin{table*}
\centering \caption{Achievable DoF and feedback overhead comparisons for the two-cell $(K+1,K,K)$ MIMO-IFBC}

\begin{tabular}{|c||c|c|c|c|}
  \hline
  \textbf{Schemes} & \textbf{Achievable} & \textbf{Channel feedback}& \textbf{Channel feedback}  &\textbf{Total channel feedback}$^{\dagger}$ \\
                   &\textbf{degrees of freedom} & \textbf{for serving cell$^{\dagger}$}& \textbf{for interfering cell$^{\dagger}$}&  \\
  \hline\hline
  CZF \cite{Kim} & $K+1$ & $2(K+1)K^{2}$ & $2(K+1)K^{2}$ & $4(K+1)K^{2}$\\ \hline
  Subspace IA \cite{Suh} & $2(K-1)$ & $2K^2$& 0 & $2K^2$ \\\hline
  \rowcolor[gray]{0.9} Proposed IA & $2K$ & $2K(K+1)$& $2(K+1)$ & $2(K+1)^2$ \\\hline
  \multicolumn{5}{l}{\scriptsize$^\dagger$The number of complex valued elements needed for channel feedback.}\\
\end{tabular}
\end{table*}
To reduce the required number of transmit antenna $M$ satisfying (10), we design the receive beamforming vectors so that $K$ effective inter-cell interference channels (IFC) are aligned within one dimensional space through the user cooperation to share CSI of interfering channels among all users in each cell. This conditions can be rewritten into more restrict conditions as follows:
\begin{eqnarray}
{\bf h}^{\mathcal{ICI}}_{i}={\bf H}^{[1,\bar{i}]\dag}_i{\bf w}^{[1,\bar{i}]}={\bf H}^{[2,\bar{i}]\dag}_i{\bf w}^{[2,\bar{i}]}=\ldots={\bf H}^{[K,\bar{i}]\dag}_i{\bf w}^{[K,\bar{i}]}.
\end{eqnarray}
The equation (11) can be represented in a matrix form, i.e.,
\begin{eqnarray}
&&\hspace{-10mm}
\underbrace{\left[%
\begin{smallmatrix}
  {\bf I}_M & -{\bf H}^{[1,\bar{i}]\dag}_i & {\bf 0} & \cdots & {\bf 0} \\
  {\bf I}_M & {\bf 0} & -{\bf H}^{[2,\bar{i}]\dag}_i & \cdots & {\bf 0} \\
  \vdots & {\bf 0} & \ddots & -{\bf H}^{[K-1,\bar{i}]\dag}_i & {\bf 0} \\
  {\bf I}_M & {\bf 0} & \cdots & {\bf 0} & -{\bf H}^{[K,\bar{i}]\dag}_i \\
\end{smallmatrix}%
\right]}_{KM\times(M+KN)}
\left[%
\begin{smallmatrix}
  {\bf h}^{\mathcal{ICI}}_i \\
  {\bf w}^{[1,\bar{i}]}\\
  {\bf w}^{[2,\bar{i}]} \\
\vdots \\
{\bf w}^{[K,\bar{i}]}
\end{smallmatrix}%
\right]\nonumber\\&&\hspace{-10mm}={\bf M}_i{\bf x}_i={\bf 0}.
\end{eqnarray}
From equation (12), we can find the receive beamforming vectors to satisfy (11) by equivalently calculating the null space of ${\bf M}_i$. It is noted that there exists a vector ${\bf x}_i$ in the null space of ${\bf M}_i$ only when the number of columns of ${\bf M}_i$ is greater than the number of rows of ${\bf M}_i$ by the number of data streams for each user, $d^{[k,\bar{i}]}$, $\forall k={1,2,\ldots, K}$, i.e.,
\begin{eqnarray}
(M+KN)-KM \geq d^{[k,\bar{i}]} = 1.
\end{eqnarray}
Moreover, if the all effective ICI channels after applying receiver beamforming vectors are aligned within a single dimension of each receiver, the minimum number of transmit antennas $M=K+1$ to remove both IUI and ICI simultaneously is necessary, i.e.,
\begin{eqnarray}
M\geq(K+1).
\end{eqnarray}
These conditions (13)-(14) now take simpler forms and can be expressed as
\begin{eqnarray}
&&\hspace{-10mm}\left(K+1\right)\leq M\leq\frac{KN-1}{K-1}\nonumber\\
&&\hspace{-15mm}\Rightarrow  K\leq \frac{(K-1)M+1}{K} \leq N.
\end{eqnarray}
To satisfy the condition (15), it is necessary to equip the numbers of transmit antennas $M=K+1$ and receive antennas $N=K$ at each BS and user at a minimum, respectively, which completes the proof.
$\,\,\,\,\,\,\,\,\,\,\,\,\,\,\,\,\,\,\,\,\,\,\,\,\,\,\,\,\,\,\,\,\,\,\,\,\,\,\,\,\,\,\,\,\,\,\,\,\,\,\,\,\,\,\,\,\,\,\,\,\,\,\,\,\,\,\,\,\,\,\,\,\,\,\,\,\,\,\,\,\,\,\,\,\,\,\,\,\,\,\,\,\,\,\,\,_{\blacksquare}$


\subsection{Discussions}
In this subsection, we compare the achievable degrees of freedom and the feedback overhead for different schemes.
For comparison, we focus on the system configuration of $(M,N,K)=(K+1,K,K)$.

\begin{itemize}
  \item \textit{\textit{\textbf{Degrees of Freedom}}}: In this case, the coordinated ZF scheme proposed by \cite{Kim} and the subspace IA scheme in \cite{Suh} can attain $K+1$ and $2(K-1)$ degrees of freedom, respectively, while our proposing scheme is able to achieve $2K$ degrees of freedom. Interestingly, in the case of $K=2$, our proposed scheme attains the optimal degrees of freedom \cite{Shin}, which coincides with the trivial outer bound in \cite{Fakhereddin}. The reason for the increase of degrees of freedom comes from that the interference alignment can utilize the signal space efficiently rather than the coordinated ZF scheme and the subspace IA do.
  \item \textit{\textit{\textbf{Feedback Overhead}}}: To compare the feedback overhead, we measure the amount of two different types of feedback information.
The proposed IA scheme needs much smaller amount of channel feedback compared to the conventional schemes with global CSI available at the BS, i.e., \cite{Kim} and \cite{Cadambe}, in the cellular networks at the expense of the user cooperation through the Wi-Fi link. For instance, the user $[1,1]$ is required to feed back the full CSI, i.e., ${\bf H}^{[1,1]}_1\in\mathbb{C}^{N\times M}$ and ${\bf H}^{[1,1]}_2\in\mathbb{C}^{N\times M}$ for performing the conventional schemes, while the proposed IA scheme only requires the effective channel vectors after applying the receive beamforming vectors explained in (8), i.e., ${\bf w}^{[1,1]\dag}{\bf H}^{[1,1]}_1\in\mathbb{C}^{1\times M}$ and ${\bf w}^{[1,1]\dag}{\bf H}^{[1,1]}_2\in\mathbb{C}^{1\times M}$. Especially, the effective interference channel vectors after ICI channel alignment for all users in the same cell are perfectly aligned along with a same direction, and this implies that only one user in each cell is enough to feed back the aligned interfering channel direction, i.e., ${\bf h}^{\mathcal{ICI}}_2$. Therefore, we can significantly diminish the total required amount of feedback information for interfering cell via uplink resources in the cellular networks with the help of user cooperation.
\end{itemize}

In conclusion, our proposed scheme has much better performance than the existing schemes while it requires much smaller amount of feedback overhead than existing schemes under full CSI assumption. The discussions above are summarized in Table I.\\

\section{Simulation results} \label{sec:Simulation Results}

Through simulation results, we first demonstrate that our proposed scheme attains the degrees of freedom analytically derived in section III and access the ergodic sum rate performance of the proposed interference alignment scheme. Throughout the simulations, the transmit power of all BSs is fixed at $P$, and noise variance of all receive antennas at a user is assumed to be the same, i.e., $\sigma^2$. The simulation results are illustrated with respect to the ratio of the total transmit power to noise variance at each receive antenna in dB scale (SNR=$\frac{P}{\sigma^2}$).

Fig. 3 illustrates the sum rate performance of the proposed interference alignment scheme according to various system configuration as a function of the SNR.
As shown in this figure, the sum rate increases linearly with the slope of 4, 8, 12, and 16 in $(3,2,2)$, $(5,4,4)$, $(7,6,6)$, and $(9,8,8)$ MIMO-IFBCs, respectively. It confirms that the sum rate performances exactly coincide with the DoF of $2K$ which is analytically derived in Section III.

To show superiority of the proposed IA scheme, we compare the sum rate performance of the proposed scheme with those of existing coordinated beamforming schemes in the $(5,4,4)$ MIMO-IFBC. As presented in Fig. 4, the proposed interference alignment method exhibits superior performance to the existing schemes described above. We also observe that the sum rate of the proposed scheme grows linearly with the slope of 8 while those of the existing schemes increase linearly with the slope of 5 and 6. From these results, we can see that the efficient utilization of the signal space leads the performance enhancement by increasing degrees of freedom.


\begin{figure}[t]
\centering
\includegraphics[width=3in]{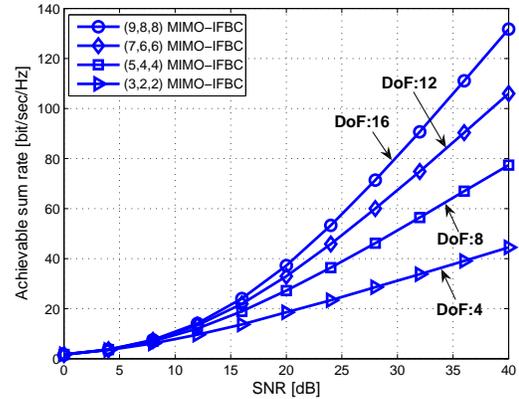}
\caption{Achievable degrees of freedom according to the number of users $K$ for $(K+1,K,K)$ MIMO-IFBC.} \label{fig_sim}
\end{figure}
\begin{figure}[t]
\centering
\includegraphics[width=3in]{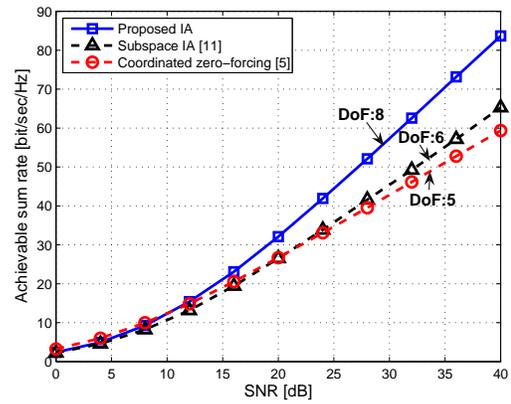}
\caption{Ergodic sum rate performance comparison for $(5,4,4)$
MIMO-IFBC.} \label{fig_sim}
\end{figure}
\section{Conclusion} \label{sec:Conclusion}
In the two-cell MIMO-IFBC, we proposed a novel cross network interference alignment algorithm exploiting benefits of the heterogeneous networks consisting of Wi-Fi systems and cellular systems. By exchanging CSI among users through Wi-Fi links, our proposed algorithm does not cause any interference to cellular links as well as significantly reduces the amount of channel feedback via cellular links. The algorithm also performs a joint design of transmit and receive beamforming vectors using the shared CSI. Furthermore, our algorithm employs intersection subspace property of the vector space to utilize signal spaces efficiently, thereby achieving higher DoF gains compared to the existing techniques. Further work will consider: i) user cooperation with limited channel feedback, and ii) the extension to the MIMO-IFBC with more than two cells.

\bibliographystyle{ieeetr}

\newpage


\end{document}